\documentstyle[twoside,fleqn,espcrc2,epsf]{article}

\newcommand{\AmS}{{\protect\the\textfont2
  A\kern-.1667em\lower.5ex\hbox{M}\kern-.125emS}}

\newlength{\numlen}

\newcommand{\nr}[1]{(\ref{#1})}
\newcommand{\fr}[2]{{\frac{#1}{#2}}}

\newcommand{\la}[1]{\label{#1}}

\newcommand{\be}{\begin{equation}}
\newcommand{\ee}{\end{equation}}
\newcommand{\ba}{\begin{eqnarray}}
\newcommand{\ea}{\end{eqnarray}}
\newcommand{\bi}{\begin{itemize}}
\newcommand{\ei}{\end{itemize}}

\newcommand{\nn}{\nonumber}

\def\lsi{\raise0.3ex
\hbox{$<$\kern-0.75em\raise-1.1ex\hbox{$\sim$}}}
\def\gsi{\raise0.3ex
\hbox{$>$\kern-0.75em\raise-1.1ex\hbox{$\sim$}}}

\newcommand{\lmax}{l_{\rm max}}

\title{%
Hard Thermal Loops and the Sphaleron Rate on the Lattice%
\thanks{Presented by K. Rummukainen
at the conference LATTICE '99, Pisa, Italy, July 1999.}%
}

\author{%
D. B{\"o}deker\address{Niels Bohr Institute, Blegdamsvej 17, DK-2100 Copenhagen \O,
Denmark},
G. D. Moore\address{Dept. of Physics,
McGill University, 3600 University St., Montreal, PQ H3A 2T8, Canada}
and
K. Rummukainen\address{NORDITA, Blegdamsvej 17, DK-2100 Copenhagen \O, Denmark}%
\hfill\raisebox{24mm}[0mm][0mm]{\makebox[0mm][r]{NORDITA-99/56HE}}%
\raisebox{19mm}[0mm][0mm]{\makebox[0mm][r]{September 1999}}
}
\begin{document}

\begin{abstract}
We measure the sphaleron rate (topological susceptibility) of hot SU(2)
gauge theory, using a lattice implementation of the hard thermal loop
(HTL) effective action.  The HTL degrees of freedom are implemented
by an expansion in spherical harmonics and truncation.  Our results
for the sphaleron rate agree with the parametric prediction of
Arnold, Son and Yaffe: $\Gamma \propto \alpha^5 T^4$.

\end{abstract}

\maketitle
\thispagestyle{empty}

\section{MOTIVATION}

Baryon number is not a conserved quantity in the Standard Model:
due to the anomaly, the violation is related to the (Minkowski time)
topological susceptibility of the SU(2) weak group.  While at low
temperatures the violation is totally negligible \cite{tHooft}, at
temperatures above the electroweak symmetry restoration temperature
($\sim 100\,$GeV) the rate of the baryon number violation ({\em
sphaleron rate}) $\Gamma$ is large.  This can have significant repercussions
for baryon number generation in the early Universe, and it opens the
avenue for purely electroweak baryogenesis.

Even though the weak coupling constant is small, at high temperatures
the sphaleron processes are dominated by IR momenta $k\sim g^2 T$ and
are thus inherently non-perturbative.  Moreover, the IR modes behave
essentially classically, which is signalled, for example, by the large
occupation numbers of the Bose fields: $n({k\sim g^2T})= (e^{k/T} -
1)^{-1}\approx T/k \sim 1/g^2 \gg 1$. This has motivated the much
utilized method of using the {\em classical equations of motion} 
to calculate $\Gamma$ in hot SU(2) theories \cite{ambjorn} (the
Higgs and fermionic degrees of freedom effectively decouple in the hot
EW phase).  For recent reviews, see \cite{Moorelat99},\,\cite{Smit97}.

The success of the classical method hinges on the efficient decoupling
of the almost-classical IR modes relevant for the sphaleron processes
and the strongly non-classical UV modes.  However, as argued by
Arnold, Son and Yaffe \cite{ASY}, this decoupling is not
complete.  A step beyond the classical approximation is the hard
thermal loop (HTL) effective theory \cite{HTL}, which incorporates the
leading order effects of the UV modes.  The HTL theory can be cast in
various forms; most practical for lattice computations is the one
where the the hard modes are described by including a large number of
classical massless particles with adjoint charge moving on the
background of IR fields.  This field + particles system can be put on
a lattice as such, and it has been succesfully used in simulations
\cite{particles}.  In this work we use an alternative Boltzmann-Vlasov
approach, where the particles are described with local density
functions $n(t, \vec x, \vec k)$.  For full description, see
\cite{bmr}.


\section{HTL THEORY ON THE LATTICE}

Let us consider a system consisting of the HTL particles moving on the
background of IR gauge fields.   The particle density functions $n(t,\vec x,
\vec k)$ obey the Vlasov equation
\ba
   \fr{{\rm d_{conv}} n}{{\rm d}t} = 0 &=& \partial_0 \delta n + 
	\vec v \cdot \vec D \delta n
 + \partial_0 \vec k \, \fr{\partial n}{\partial \vec k} \nn\\
    &=& v\cdot D \delta n + g  v_i F^{0i} \, 
     \fr{\partial n_0}{\partial k} \,, \la{vlasov}
\ea
where $n_0 = (e^{k/T} - 1)^{-1}$, $n = n_0 + \delta n^a$, and the
Lorentz-force $\propto \vec v \times \vec B$ has been neglected.  The
IR gauge fields evolve according to the Yang-Mills equations:
\be
    D_\mu F^{\mu\nu} = j^\nu_{\rm hard} = 
    4 g \int \fr{d^3 k}{(2\pi)^3}\,\, v^\nu \delta n,  \la{ym}
\ee
where the 4-velocity $v = (1,\vec k/k)$.  These equations can be
further simplified by factorizing $\delta n^a = -g W^a(x,\vec
v)(\partial n_0/\partial k)$ and integrating over the amplitude $|\vec
k|$ \cite{BlaizotIancu}:
\ba
  D_\mu F^{\mu\nu} &=& m_D^2 \int \fr{d\Omega}{4\pi} \, v^\nu W(x,\vec v) \nn \\
  v^\mu D_\mu \, W(x,\vec v) &=&  v_i F^{0i}  \la{BI}
\ea
Here $d\Omega$ integration is over the directions of the 4-velocity
$v$.  The field $W^a(x,\vec v)$ is proportional to the flux of the
particles at point $x$ to direction $\vec v$.

In order to perform lattice simulations the field $W$ has to be
regularized in space (standard lattice) and on the $\vec v$-sphere.
We do this by expanding $W$ in spherical harmonics: $W^a(x,\vec v) =
W_{lm}^a(x) Y_{lm}(\vec v)$, and truncating the expansion to $l \le
\lmax$.  In terms of $W_{lm}$, the equations \nr{BI} finally become \cite{bmr}
\ba
   D_i F^{i0} &=& (m_D^2/\sqrt{4\pi}) W_{00}  \la{gauss}\\
   D_\mu F^{\mu i} &=& (m_D^2/4\pi) V_m^{i\,*} W_{1m} \la{lmeq} \\
   D_0 W_{lm} &=& -C_{lm;i}^{l'm'} \, D_i W_{l'm'} + \delta_{l,1} V_m^i F^{0i} \la{lmeq2}\,.
\ea
Here the coefficients $C_{lm;i}^{l'm'} = \int d\Omega Y_{lm}^* v^i
Y_{l'm'}$ and $V_m^i = \int d\Omega Y_{1m} v^i$.  Eq.~\nr{gauss} is
the Gauss law, and, as long as it is satisfied by the initial
configuration, it is preserved by Eqs.~\nr{lmeq} and \nr{lmeq2}.

With a finite $\lmax$, these equations can be readily discretized:
SU(2) gauge field is defined on the links of the lattice, and the
$(\lmax+1)^2$ adjoint $W^a_{lm}$ fields are on lattice sites.  The
discretization and the properties of the theory on the lattice are
discussed in detail in \cite{bmr}.

\section{THE SPHALERON RATE}

The measurement of the sphaleron rate proceeds along similar lines to
the purely classical theory: first, we generate an ensemble of initial
thermalized configurations (which satisfy the Gauss law), and then
evolve these with Eqs.~\nr{lmeq},\nr{lmeq2}.  We then obtain $\Gamma$
by measuring the rate of the Chern-Simons number diffusion
\cite{bmr,ambjorn,Moorelat99}.  

We have to check how $\Gamma$ depends on (a) $\lmax$, (b) lattice
spacing and (c) the Debye mass $m_D$.  Only the last parameter is
physical (it depends on the particle content of the theory).

Let us first consider the $\lmax$ dependence.  In Fig.~\ref{fig:lmax} we
show $\Gamma$ measured from a set of lattices with $\lmax \le 10$.  We
note that when $\lmax$ is even, the rate remains remarkably constant
(much better than indicated by naive arguments \cite{bmr}).  The
behaviour at odd $\lmax$ can be understood by considering the
properties of the gauge field propagator \cite{bmr}.  Thus, we
conclude that modest values of $\lmax \approx 4$--6 are sufficient in
order to obtain the $\lmax\rightarrow\infty$ behaviour within
reasonable statistical errors.

\begin{figure}[bt]

\epsfxsize=6.3cm\epsfbox{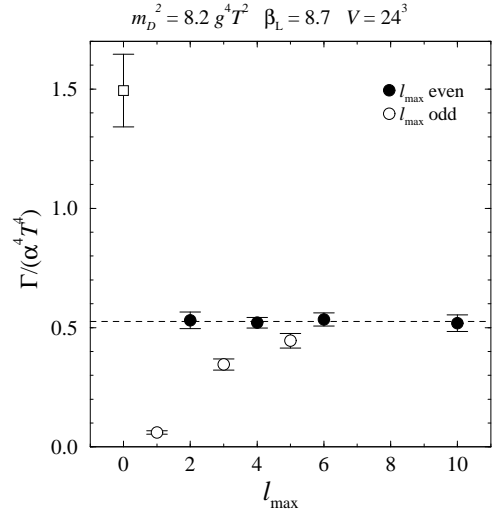}

\vspace*{-8mm}
\caption[a]{
The sphaleron rate dependence on the $l_{\rm max}$ cutoff.}
\la{fig:lmax}
\end{figure}




\begin{figure}[bt]

\epsfxsize=6.4cm\epsfbox{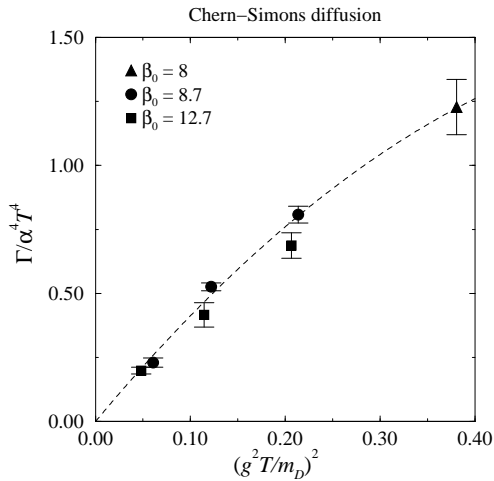}

\vspace*{-10mm}
\caption[a]{
The sphaleron rate $\Gamma$ in HTL theory.  In leading order $\beta_0
= 4/g^2 a$, where $a$ is the lattice spacing.}
\la{fig:kappa}
\vspace*{-6mm}
\end{figure}

Dimensionally, one would expect that $\Gamma \propto \alpha^4 T^4$, the
non-perturbative scale to the fourth power.  However, as argued by
Arnold, Son and Yaffe \cite{ASY}, the evolution of the IR fields is
Landau damped by the UV modes, and the rate is slower by one further
factor of $\alpha$, parametrically
\be
  \Gamma = \kappa' \fr{g^2T^2}{m_D^2} \, \alpha^5 T^4\,,  \la{kappa}
\ee
where $\kappa'$ is a constant to be determined by lattice
measurements.  In Fig.~\ref{fig:kappa} we show the behaviour of
$\Gamma$ against $g^2T^2/m_D^2$, measured using various lattice
spacings $a \propto 1/\beta_0$.  The rate is clearly not constant when
the Debye mass is varied, and it goes to zero when
$m^2_D\rightarrow\infty$, as predicted by Eq.~\nr{kappa}.  We have not
observed any significant dependence on the lattice spacing (provided
that it is small enough).  It should be noted that the physical
$m_D^2$ is not only the `bare' $m_D^2$ which appears in Eq.~\nr{lmeq},
but it is a sum of the bare $m_D^2$ and a contribution due to the
UV lattice gauge field modes $\propto 1/a$ \cite{Arnoldlatt}.

\begin{figure}[bt]

\epsfxsize=6.3cm\epsfbox{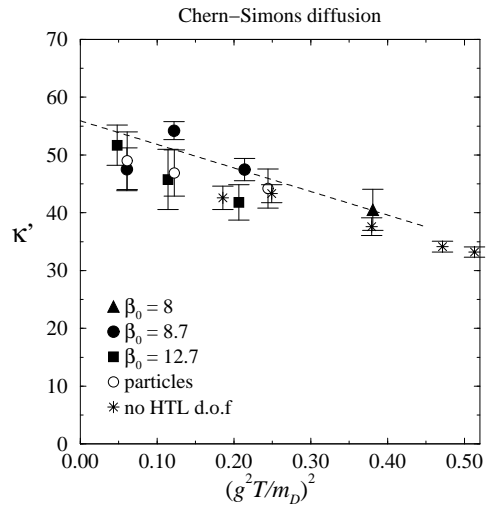}

\vspace*{-8mm}
\caption[a]{
Comparison of $\kappa'$, as measured from different simulations.
Filled symbols: this work; open circles: `particles' method
\cite{particles}; bursts: classical SU(2) gauge theory simulation \cite{mr}.}
\la{fig:kappaprime}
\end{figure}

In Fig.~\ref{fig:kappaprime} we compare the coefficient $\kappa'$ of
the scaling law \nr{kappa} as measured in this work, with the
particles method \cite{particles}, and with only the classical SU(2)
gauge theory evolution without any added HTL degrees of freedom
\cite{mr}.  In the last case the physical $m_D^2$ arises solely
through the lattice UV modes; here the lattice spacing is up to a
factor of $\sim$ 4 smaller than in the two HTL approaches.  The consistency
of the results is remarkable, considering the very different treatments
used.

To conclude with, the sphaleron rate in hot SU(2) gauge theory is now
settled.  Inserting the Standard Model value of $m_D^2 = 11/6\,g^2
T^2$, we obtain for the rate a value $\Gamma = (25\pm 2) \alpha^5
T^4$.

\end{document}